\documentclass[modern]{aastex631}

\usepackage{amsmath}

\newcommand{\gaia}{{\it Gaia}}

\newcommand{\todo}[1]{\textcolor{black}{#1}}

\begin{document}

\title{Where are {\it Gaia}'s small black holes?}
\author[0000-0002-1980-5293]{Maya Fishbach}
\affiliation{Canadian Institute for Theoretical Astrophysics, 60 St George St, University of Toronto, Toronto, ON M5S 3H8, Canada}
\affiliation{David A. Dunlap Department of Astronomy and Astrophysics, University of Toronto, 50 St George St, Toronto ON M5S 3H4, Canada}
\affiliation{Department of Physics, 60 St George St, University of Toronto, Toronto, ON M5S 3H8, Canada}

\author[0000-0001-5228-6598]{Katelyn Breivik}
\affiliation{McWilliams Center for Cosmology \& Astrophysics, Department of Physics, Carnegie Mellon University, Pittsburgh, PA 15213, USA}

\author[0000-0003-1817-3586]{Reinhold Willcox}
\affiliation{Institute of Astronomy, KU Leuven, Celestijnenlaan 200D, 3001 Leuven, Belgium} 
\affiliation{Leuven Gravity Institute, KU Leuven, Celestijnenlaan 200D, box 2415, 3001 Leuven, Belgium }

\author[0000-0001-5484-4987]{L.~A.~C.~van~Son}
\affiliation{Center for Computational Astrophysics, Flatiron Institute, 162 Fifth Avenue, New York, NY 10010, USA}
\affiliation{Department of Astrophysical Sciences, Princeton University, 4 Ivy Lane, Princeton, NJ 08544, USA}
\affiliation{Department of Astrophysics/IMAPP, Radboud University, P.O. Box 9010, NL-6500 GL Nijmegen, The Netherlands}

\begin{abstract}
{\it Gaia} has recently revealed a population of over 20 compact objects in wide astrometric binaries, while LIGO-Virgo-KAGRA (LVK) have observed around 100 compact object binaries as gravitational-wave (GW) mergers. 
Despite belonging to different systems, the compact objects discovered by both {\it Gaia} and the LVK follow a multimodal mass distribution, with a global maximum at neutron star (NS) masses ($\sim 1$--$2\,M_\odot$) and a secondary local maximum at black hole (BH) masses $\sim10\,M_\odot$.
However, the relative dearth of objects, or ``mass gap," between these modes is more pronounced among the wide binaries observed by {\it Gaia} compared to the GW population, with $9^{+10}_{-6}\%$ of GW component masses falling between $2.5$--$5\,M_\odot$ compared to $\lesssim5\%$ of {\it Gaia} compact objects. 
We explore whether this discrepancy can be explained by the natal kicks received by low-mass BHs.
GW progenitor binaries may be more likely to survive natal kicks, because the newborn BH has a more massive companion and/or is in a tighter binary than {\it Gaia} progenitor binaries.
We compare the survival probabilities of {\it Gaia} and GW progenitor binaries as a function of natal kick strength and pre-supernova binary parameters, and map out the parameter space and kick strength required to disrupt the progenitor binaries leading to low-mass BHs in {\it Gaia} systems more frequently than those in GW systems.
\end{abstract}

\section{Introduction}
\label{sec:intro}
New observations across the electromagnetic and gravitational-wave spectra are rapidly expanding our knowledge of the stellar graveyard. 
With novel observatories and detection methods, the neutron star (NS) and black hole (BH) remnants of massive stars can now be observed in a variety of different systems, including as isolated objects~\citep{2022ApJ...933...83S,2022ApJ...933L..23L}, in wide binaries with luminous companions~\citep[e.g.,][]{2022NatAs...6.1085S,2023MNRAS.518.1057E,2023MNRAS.521.4323E,2024OJAp....7E..58E, 2023AJ....166....6C,2024A&A...686L...2G,2025arXiv250523151A}, in tight binaries accreting from their stellar companions~\citep{1972NPhS..240..124B,2006csxs.book..623T,2006AAS...209.0705R,2014SSRv..183..223C, 2015PhR...548....1M}, and in gravitational-wave (GW) driven binary mergers with other compact objects~\citep{2016PhRvL.116f1102A,2019PhRvX...9c1040A,2021PhRvX..11b1053A,2023PhRvX..13d1039A,2020PhRvD.101h3030V,2021ApJ...922...76N}.
By comparing these different compact object populations, we can probe the shared as well as the divergent aspects of their evolutionary histories~\citep[e.g.][]{2022ApJ...929L..26F,2022ApJ...938L..19G,2023ApJ...946....4L}. 

In this work, we focus on comparing the population of NSs and BHs in wide astrometric binaries discovered with \gaia{}~\citep{2016A&A...595A...1G} to the population of GW binary mergers discovered by the LIGO-Virgo-KAGRA (LVK) GW observatory~\citep{2015CQGra..32g4001L,2015CQGra..32b4001A,2021PTEP.2021eA101A}. Analyses of \gaia's third data release~\citep[DR3;][]{2023A&A...674A...1G} coupled with spectroscopic followup resulted in 21 NS candidates~\citep{2024OJAp....7E..58E}, two $\sim10\,M_\odot$ BHs~\citep{2023AJ....166....6C,2023MNRAS.518.1057E,2023MNRAS.521.4323E} and a $\sim3.6\,M_\odot$ ``mass gap" BH candidate, G3425~\citep{2024NatAs...8.1583W}.
Another $\sim30\,M_\odot$ BH, BH3, was discovered in pre-release DR4 observations~\citep{2024A&A...686L...2G}, and the full DR4 is expected to contain $\mathcal{O}(100)$ more compact object discoveries~\citep{2022ApJ...931..107C,2023A&A...670A..79J,2024NewAR..9801694E,Nagarajan+2025:2025PASP..137d4202N}. Additional detections of BHs in wide binaries are possible with radial velocity surveys, including the recent $\approx10\,M_\odot$ BH candidate in a wide binary with an $\approx11\,M_\odot$ Be star companion~\citep{2025arXiv250523151A}.

Meanwhile, the recent GW catalog from the LVK collaboration, GWTC-3, contains 70 confident binary black hole (BBH) observations, two binary neutron star (BNS) systems, and four neutron star--black holes (NSBHs), adopting the detection thresholds from~\citet{2023PhRvX..13a1048A} and counting GW190814, with a secondary component weighing $2.6\,M_\odot$, as a BBH. An additional NSBH event from the beginning of the LVK's ongoing fourth observing run (O4) has been reported~\citep[GW230529,][]{2024ApJ...970L..34A}, and the ongoing fourth observing run O4 is expected to yield hundreds of additional compact binary mergers~\citep{2018LRR....21....3A}.

At a first glance, the low-mass end of the compact object mass distribution observed by the LVK and \gaia{} appear remarkably similar, with a broad NS mass distribution extending above the Chandrasekhar limit~\citep{2020ApJ...892L...3A,2024OJAp....7E..58E,Schiebelbein}, an abundance of $\sim10\,M_\odot$ BHs~\citep{2023ApJ...946...16E,2023PhRvX..13a1048A,2023ApJ...955..107F,2024PhRvX..14b1005C}, and a relative dearth of masses in between~\citep{2022ApJ...931..108F,2022ApJ...937...73Y,2023MNRAS.518.5298B,2023PhRvX..13a1048A}.\footnote{\todo{\gaia's BH3 suggestively overlaps with another overdensity in the BBH mass distribution at $\sim33\,M_\odot$~\citep{2023PhRvX..13a1048A, 2024ApJ...962...69F}, but because its selection for early publication in advance of DR4 introduces a poorly understood selection effect, we limit our comparison to the low-mass end of the BH mass distribution in this work.}}
This dearth of low-mass BHs with masses between the heaviest NS and $\sim10\,M_\odot$ is reminiscent of the purported ``mass gap" first observed in X-ray binaries~\citep{1998ApJ...499..367B,2010ApJ...725.1918O,2011ApJ...741..103F}, whose origin, whether physical~\citep{2012ApJ...749...91F} or an observational selection effect~\citep{2001ApJ...554..548F,2012ApJ...757...36K,2023ApJ...954..212S}, has long been debated. Meanwhile, GW observations of systems such as GW190814 and GW230529, in which one of the binary components appears to be squarely in the gap, have proven that the mass gap is not empty, at least among GW binaries~\citep{2020ApJ...896L..44A,2022ApJ...931..108F,2023PhRvX..13a1048A,2024ApJ...970L..34A}.
If G3425 is indeed a $3\,M_\odot$ BH, it would also prove that \gaia{} wide binaries can occupy the mass gap~\citep{2024NatAs...8.1583W}. 
In other words, for the GW and \gaia{} compact object populations, the ``gap" may be better described as a ``dip."
In the remainder of this work, we use the term ``gap" interchangeably with ``dip" or ``dearth" in order to encompass the possibility of a non-empty gap.

There are several proposed explanations for the NS--BH mass gap, each of which predicts different gap properties (i.e. existence and depth) among the compact objects in different systems. The mass gap may originate from the supernova remnant mass function~\citep{2001ApJ...554..548F,2020MNRAS.499.3214M}, in which case all compact objects of stellar origin would universally share the same mass gap. On the other hand, the mass gap may depend on variables like metallicity or mass transfer history, in which case systems that experience divergent evolutionary pathways, like GW versus \gaia{} versus X-ray binaries, need not share similar gap properties~\citep{2022ApJ...940..184V}. 

In the context of isolated binary evolution, supernova natal kicks are a proposed explanation for the existence of a NS--BH mass gap among compact objects in binary systems~\citep[e.g.][]{2001ApJ...554..548F}. 
While the kicks imparted on newborn BHs are highly uncertain, if low-mass BHs tend to receive large natal kicks (e.g. `fixed momentum' kicks from supernova fallback, implying a kick velocity that scales inversely with BH mass; \citealt{2012ApJ...749...91F}) or be accompanied by significant mass loss (leading to a Blaauw kick;~\citealt{1961BAN....15..265B}), their birth may preferentially disrupt their binary~\citep{2001ApJ...554..548F}.
This may lead to a relative dearth of low-mass BHs in binary systems, compared to NSs, some of which are thought to receive small kicks $\mathcal{O}(10)$ km/s~\citep{2004ApJ...612.1044P,2025arXiv250508857V}, and $\gtrsim 10\,M_\odot$ BHs, which are predicted to receive the smallest kicks on average~\citep{2020MNRAS.499.3214M,2023arXiv231112109B,2024Ap&SS.369...80J}, although the distribution of kicks as a function of remnant mass is highly uncertain; see further discussion in \S\ref{subsec:kick-background}.
Because the outcome of the kick depends on the initial binary parameters, the properties of any resulting mass gap would vary among different compact object populations (i.e. \gaia{} versus GW binaries). Binaries that are more tightly bound prior to the birth of the low-mass BH, due to higher companion masses or shorter orbital periods, would more likely survive the natal kick, creating an evolutionary selection bias that favors the presence of low-mass BHs in certain binary systems over others.

As a first step towards disentangling whether the NS-BH mass gap is universal (an imprint of the supernova remnant mass function) or evolutionary (e.g., dependent on mass transfer or natal kicks), we compare the statistics of NS and low-mass BH masses founds in \gaia{} versus GW binaries. We quantify the discrepancy in the depth of the NS-BH mass gap between \gaia{} and GW systems, finding that with current data, the mass gap appears emptier for \gaia's NSs and BHs (\S\ref{sec:massgapstatistics}). We investigate this discrepancy in the context of BH kicks, comparing the survival probabilities between different binary systems and discussing implications for low-mass BH natal kicks, supernova mass loss, and pre-supernova binary properties (\S\ref{sec:kicks}). We conclude by summarizing our results (\S\ref{sec:conclusion}).

\section{Gaia's NS-BH mass gap compared to GW binaries}
\label{sec:massgapstatistics}

\begin{figure}
\centering
    \includegraphics[width=0.5\textwidth]{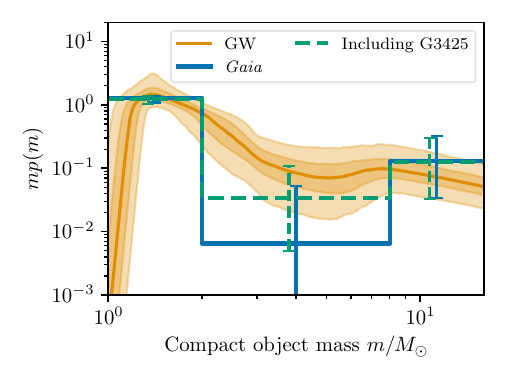}
    \caption{Mass distribution of NSs and low-mass BHs ($m < 16\,M_\odot$) found in GW merging binaries (orange) and \gaia{} wide binaries (blue). The GW component mass distribution, \todo{which includes both compact objects in each binary}, is the \textsc{Power Law + Break + Dip} model fit to GWTC-3 plus the event GW230529 from O4 (but is consistent with the GWTC-3-only fit) from \citet{2024ApJ...970L..34A}. The \gaia{} mass distribution is a piecewise constant model fit to a sample of 21 NSs, BH1 and BH2 (blue, solid line). Alternatively, we include the \gaia{} $\approx3.6\,M_\odot$ BH candidate G3425 (dashed green). The NS and low-mass BHs in both \gaia{} and GW systems share a remarkably similar distribution, but the ``mass gap" region seems emptier for \gaia{}. \label{fig:Gaia_versus_LVK_massdist}}
\end{figure}

The mass distribution of the NSs and low-mass BHs (below $16\,M_\odot$) found in \gaia{} wide binaries and GW sources is shown in Fig.~\ref{fig:Gaia_versus_LVK_massdist}. 
In orange, we show the distribution of GW component masses (including both masses in each GW binary) inferred in~\citet{2024ApJ...970L..34A}, fit to the phenomenological \textsc{power law + dip + break} mass model introduced by~\citet{2020ApJ...899L...8F} and \citet{2022ApJ...931..108F}. \citet{2024ApJ...970L..34A} also presents consistent results inferred using the semi-parametric \textsc{binned Gaussian process} model~\citep{2017MNRAS.465.3254M,2020ApJ...891L..31F,2023ApJ...957...37R}; see their Fig. 6.
Although this mass distribution includes the ``mass-gap" NSBH event GW230529 from the LVK's O4, this single event does not noticeably affect the depth of the mass gap, and the resulting mass distribution is consistent with the pre-O4 GWTC-3 inference~\citep{2023PhRvX..13a1048A}.
\todo{The GW mass distribution shown in Fig.~\ref{fig:Gaia_versus_LVK_massdist} includes both component masses in each binary, but the inferred distribution of primary masses (the heavier of the two components) is nearly indistinguishable. This is largely a consequence of the modeling prior, because \textsc{Power law + break + dip} fits for an underlying component mass distribution from which primaries and secondaries are both drawn~\citep[see discussion in][]{2020ApJ...891L..27F,2024ApJ...962...69F}. There are insufficient low-mass GW events to determine whether the primary and secondary mass distributions differ. We choose to include both component masses in our comparison to \gaia{} to remain agnostic about which component is the first born compact object in the binary.} 

In solid blue, we show a histogram fit to \gaia{}'s NS and BH masses, using the 21 NSs reported in \citet{2024OJAp....7E..58E} and \gaia{} BH1 and BH2. Motivated by the structure in the GW mass distribution, we choose three bins for this figure: 1--$2\,M_\odot$, 2--$8\,M_\odot$ and 8--$16\,M_\odot$, so that the first bin contains all of the NSs, the second bin contains any ``mass gap" objects (of which there are none for the version shown in blue) and the third bin contains BH1 and BH2. (The measurement uncertainties in the \gaia{} compact object masses are sufficiently small that we neglect ambiguity in bin assignment.) \todo{While the second ``mass gap" bin is twice as large (in log-space) as the ``NS" and ``BH" bins, this bin contains at most one \gaia{} system, so there is not enough information to probe smaller-scale structure. The ``NS" bin contains enough events to probe the \gaia{} NS mass distribution in finer resolution, but this is beyond the scope of this work~\citep[see][]{Schiebelbein}.} Assuming a Poisson likelihood for the observed number of systems $N_i$ in each bin $i$ given expected number $R_i$, and placing an independent Jeffreys prior\footnote{A Jeffreys prior is a convenient choice because it is non-informative prior that is invariant under reparameterizations (e.g. if we were to infer $\log R_i$ rather than $R_i$).} on each $R_i$ (i.e., $p(R_i) \propto 1/\sqrt{R_i}$), the posterior probability distribution is:
\begin{equation}
\label{eq:binheightposterior}
    p(R_i \mid N_i) \propto R_i^{N_i - 0.5} e^{-R_i}.
\end{equation}
While the GW mass distribution is corrected for observational selection effects (i.e. the observational bias towards detecting high-mass GW binaries), we do not fold in selection effects for the \gaia{} mass distribution, because these are thought to depend only weakly on compact object mass~\citep{2024arXiv241100654L}.
However, the \gaia{} DR3 selection effects on orbital period are significant~\citep{2023A&A...674A..25H}, as we discuss further in \S\ref{sec:conclusion}.
The mass distributions shown in Fig.~\ref{fig:Gaia_versus_LVK_massdist} are normalized probability densities over the range 1--$16\,M_\odot$, so that the height of the histogram shows the posterior median on the relative rate in each bin, and the error bars represent the symmetric 90\% credibility intervals.
In addition to the blue histogram, we show the \gaia{} mass distribution including the $\sim3.6\,M_\odot$ \gaia{} BH candidate G3425~\citep{2024NatAs...8.1583W} in dashed green. This increases the relative rate in the central ``mass gap" bin.
(We consider G3425 separately here because it was identified by an independent team compared to the 2 BHs and 21 NSs included in the blue histogram, and therefore decreases the uniformity of the sample.) 
We also checked the sensitivity of the \gaia{} mass distribution to the number of NSs, because it is possible that the NS sample from \citet{2024OJAp....7E..58E} is contaminated with white dwarfs. Even under the extreme assumption that half of the NSs from \citet{2024OJAp....7E..58E} are white dwarfs, the resulting mass distribution, particularly the inferred fraction of sources in the ``mass gap" bin, is consistent within uncertainties. 

The \gaia{} and GW mass distributions in Fig.~\ref{fig:Gaia_versus_LVK_massdist} are intriguingly consistent with each other. Both populations display a ``bimodality"~\citep{2024OJAp....7E..58E}, with the first mode at NS masses between $1$--$2\,M_\odot$ and the second mode at $\sim10\,M_\odot$. 
The GW mass distribution shown here parametrizes this bimodality in terms of a broken power law with a dip, and recovers a global maximum at $1.28^{+0.15}_{-0.17}\,M_\odot$ and a secondary local maximum at $6.4^{+1.9}_{-2.5}
\,M_\odot$ (this secondary local maximum is exhibited in 97\% of posterior traces). Here and throughout, we report medians and 90\% credibility intervals.
We note that the GW population model shown here does not explicitly allow for a peak in the low-mass BBH region. Other, more flexible models have found evidence for a pronounced overdensity at BBH component masses of $\approx10\,M_\odot$~\citep{2021ApJ...913L..19T,2023ApJ...946...16E,2023PhRvX..13a1048A,2024PhRvX..14b1005C}; see~\citet{2023ApJ...955..107F} for a discussion of the difference between the $\approx10\,M_\odot$ peak and the $\approx6\,M_\odot$ ``turn-on" of the BBH mass distribution. 
While the $10\,M_\odot$ peak in the BBH mass distribution makes the comparison with \gaia{} even more striking (both BH1 and BH2 are $\approx9\,M_\odot$, placing them in the peak), the focus of this work is on the transition between NS and BH masses rather than on substructure within the BH mass spectrum. Allowing for such substructure does not change the relative rates of ``mass gap" versus ``above gap" BBHs. 

\begin{figure*}
    \includegraphics[width=\textwidth]{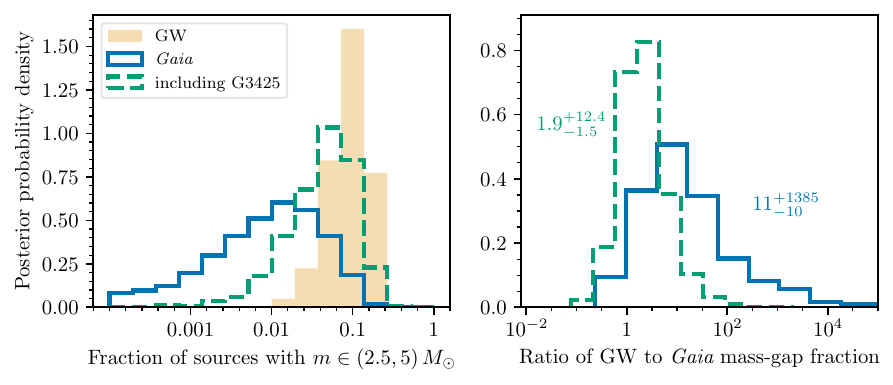}
    \caption{{\em Left}: Posterior probability densities of the fraction of ``mass gap" BHs with mass in the range 2.5--5$\,M_\odot$ among GW binaries (orange filled histogram), the \gaia{} standard sample (blue solid line) or the \gaia{} sample including the mass-gap BH G3425 (green dashed line).
    {\em Right}: Ratio of the GW mass-gap fraction (reported in the orange, filled histogram on the left) to the {\it Gaia} mass-gap fraction excluding (blue, solid histogram) or including (green, dashed histogram) G3425.
    Mass-gap BHs weighing $\approx3\,M_\odot$ are relatively more common in GW binaries than in \gaia{} binaries, although including G3425 in the \gaia{} sample reduces the tension. \label{fig:ratio_MG_rate}}
\end{figure*}

A consequence of the bimodality is the relative dearth of low-mass black holes in the ``mass gap". While the GW and \gaia{} populations both exhibit this relative dearth, it seems more pronounced for the BHs in \gaia{} binaries compared to those in GW binaries. To make this comparison concrete, the left panel of Fig.~\ref{fig:ratio_MG_rate} shows the fraction of compact objects with mass between 2.5--5$\,M_\odot$ among GW binaries with component masses $m<16\,M_\odot$\footnote{While we focus on the low-mass end of the GW mass distribution for the comparison with \gaia{}, this choice does not significantly affect our results, because the range $m<16\,M_\odot$ includes $97^{+2}_{-4}\%$ of component masses in GW binaries, and $94^{+3}_{-7}\%$ of GW component masses are below $10\,M_\odot$.} (orange, filled histogram), compared to those in \gaia{} binaries (with and without the mass-gap event G3425, in solid blue and dashed green, respectively). 
The fraction of GW sources in this mass range is calculated from integrating the mass distribution of Fig.~\ref{fig:Gaia_versus_LVK_massdist} (in orange) over the $(2.5, 5)\,M_\odot$ interval. 
For the \gaia{} mass-gap fraction, we calculate the rate in the specified mass range according to Eq.~\ref{eq:binheightposterior} and divide by the sum of inferred rates across all bins.
\todo{We choose a restricted ``mass gap" range 2.5--5$\,M_\odot$ as opposed to the broader $2$--$8\,M_\odot$ range for this comparison as a conservative choice, to avoid inflating the difference between the GW and \gaia{} ``mass gap" fractions. A wider range (e.g., 2--$8\,M_\odot$) corresponds to a higher GW ``mass gap" rate by a factor $>2$, but an unchanged \gaia{} rate, because the \gaia{} ``mass gap" bin contains the same number of sources for either definition.}
The fraction of compact objects in the mass gap is probably larger among GW binaries than \gaia{} binaries, with credibility 94\% (or 76\% if including G3425). 

In fact, the right panel of Fig.~\ref{fig:ratio_MG_rate} shows that mass-gap BHs are $11^{+1385}_{-10}$ times more common among GW systems than \gaia{} systems, or $1.9^{+12.4}_{-1.5}$ times more common if G3425 is included in the \gaia{} sample (90\% credibility). 
This is in contrast to the fraction of sources with masses in the 1--2$\,M_\odot$ range or 8--16$\,M_\odot$ range, which are notably consistent between the GW and $\gaia{}$ samples, as seen in Fig.~\ref{fig:Gaia_versus_LVK_massdist}.

Based on the higher fraction of mass-gap BHs among GW systems compared to \gaia{} systems, we adopt the hypothesis that the dearth of low-mass BHs in the {\it Gaia} sample relative to the GW sample is influenced by the evolutionary histories of these systems. 
Making the simplifying assumption that both populations originate from binary stellar evolution (but see further discussion in \S\ref{sec:conclusion}), we investigate how differences in their pre-supernova orbital architectures may lead to the observed difference in the mass gap.


\section{Do kicks disrupt binaries with low-mass black holes?}
\label{sec:kicks}

\subsection{Expectations for NS and BH kicks}
\label{subsec:kick-background}

Several processes in single and binary star evolution shape the compact object mass distribution and may impact the relative rates of mass-gap black holes in different systems. These factors include stellar mass loss (and its dependence on metallicity), binary mass transfer, supernova mass loss and supernova natal kicks. 
For example, \citet{2022ApJ...940..184V} suggested that the stable mass transfer channel is responsible for the dearth of merging BBH systems with component masses below $\approx9\,M_\odot$. This is an example of an evolutionary selection effect acting on BHs in GW merging binaries, but not necessarily BHs in other systems. Similarly, in low-mass X-ray binaries, accretion onto a heavy NS can create a mass-gap BH, unlike in systems that do not experience long-lived, stable accretion~\citep{2022ApJ...931..107C,2023ApJ...954..212S}. 
It may also be that stellar stripping by a binary companion affects the supernova properties, including the resulting compact object mass~\citep{2018MNRAS.479.3675M,2021ApJ...916L...5V}. 

Another example of an evolutionary selection effect is NS and BH kicks, including natal kicks caused by supernova asymmetries and Blaauw kicks caused by supernova mass loss.
From supernova theory, we expect that low-mass NSs tend to receive smaller natal kicks ($\lesssim100$ km/s) than high-mass NSs, while the trend is reversed for BHs, with the highest mass BHs receiving no natal kicks in a direct collapse (non-explosive) supernova~\citep{2012ApJ...749...91F, 2020MNRAS.499.3214M,2024Ap&SS.369...80J,2023arXiv231112109B}. 
Supernova simulations have suggested that the lowest mass (i.e. ``mass gap") BHs may receive even larger natal kicks than many NSs, reaching hundreds and possibly even up to 1000 km/s~\citep{Janka+2013:2013MNRAS.434.1355J,2023arXiv231112109B}.

Observationally, supernova kicks are encoded in binary properties, such as the peculiar velocity and orbital parameters for \gaia{} and other non-interacting binaries~\citep[e.g.][]{2023ApJ...959..106B,2024MNRAS.535.3577K,2024PhRvL.132s1403V,2025PASP..137c4203N,2025arXiv250416669W} and the spin tilts for GW binaries (e.g.~\citealt{2000ApJ...541..319K,2016ApJ...832L...2R,2017PhRvL.119a1101O,2021ApJ...920..157C}, but see also \citealt{2024arXiv241203461B}). For example, the eccentricities of \gaia's NS and BH sample are consistent with initially (pre-supernova) circular binaries that receive nonzero $\mathcal{O}(10-100)$ km/s natal kicks~\citep{2023MNRAS.518.1057E,2023MNRAS.521.4323E,2024OJAp....7E..58E}. 
NSs and BHs in other systems, such as X-ray binaries and isolated compact objects, provide complementary constraints~\citep[e.g.][]{2023MNRAS.525.1498Z,2022ApJ...930..159A}.
Recent work on Be X-ray binaries—systems containing a neutron star and a rapidly rotating companion—finds evidence for a neutron star kick distribution with both a very low-velocity component ($<10\,\mathrm{km/s}$) and a higher-velocity component around $100\,\mathrm{km/s}$ \citep{2025arXiv250508857V}. 
Isolated young pulsars with measured velocities also provide an estimate of NS kicks, favoring a lognormal distribution that peaks at $\approx150$--$200$ km/s~\citep{2025arXiv250522102D}.
For a compilation of BH kick measurements (in binary systems), see, e.g., Fig. 6 in~\citet{{2025ApJ...979..209V}}. 

It remains unclear whether the observations support the theoretical kick predictions described above.
For example, the \gaia{} mass-gap BH candidate G3425 must have received a small natal kick ($\lesssim10$ km/s) because of its wide and circular orbit~\citep{2024NatAs...8.1583W}.
From both simulations and observations, it seems likely that there is a wide spread of kick velocities, possibly spanning hundreds of km/s, even at a fixed remnant mass~\citep{2020MNRAS.499.3214M,2025PASP..137c4203N}.

Of course, observations of NSs and BHs in binary systems are biased tracers of kick magnitudes and orientations because the binary must have survived the kick~\citep{1996ApJ...471..352K}. 
Although there may be exceptions (cases in which certain binaries may only exist because of natal kicks that are fortuitously oriented opposite to the velocity vector of the exploding star; e.g. \citealt{2000ApJ...541..319K,2023ApJ...953..152O}), kicks generally tend to disrupt binaries~\citep{2019A&A...624A..66R}. This may lead to an under-representation of compact objects (potentially including low-mass black holes) that receive large kicks in binary systems. 
In the following, we assume natal kicks are isotropically oriented, but they may instead be preferentially oriented parallel to the orbital plane (equatorial) or perpendicular to the orbital plane (polar). In fact, recent observational evidence from Be X-ray binaries favors polar NS kicks~\citep{2025arXiv250508857V}, supporting earlier results from isolated pulsars~\citep{2006ApJ...639.1007W}. {At a fixed kick magnitude, preferentially polar natal kicks would lead to lower survival probabilities than the isotropic kicks we consider here~\citep{2000ApJ...541..319K}.} 

\subsection{Ratio of survival probabilities between different orbital configurations}

\begin{figure*}
\includegraphics[width=0.9\textwidth]{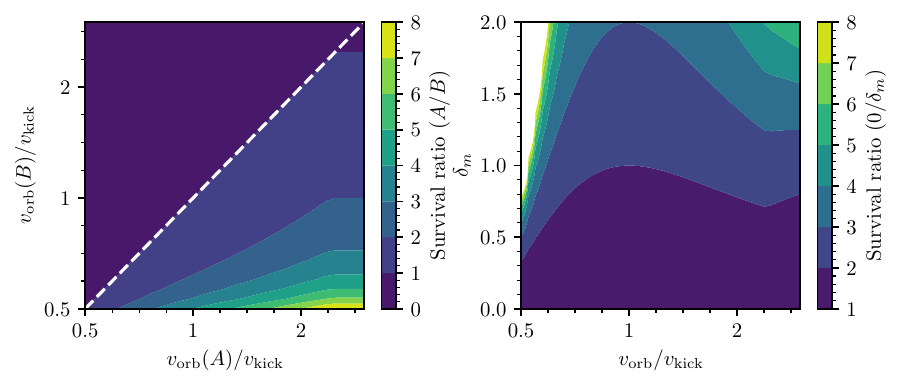}
\caption{\label{fig:ratio_survival_prob_contours_dimensionless}
Ratio of post-supernova binary survival probabilities between two binaries with (a) {\it left panel}: negligible supernova mass loss, but different pre-supernova orbital velocities relative to the natal kick $v_\mathrm{orb}/v_\mathrm{kick}$; (b) {\it right panel}: the same $v_\mathrm{orb}/v_\mathrm{kick}$, but one binary has negligible supernova mass loss $\delta_m = 0$ and the other has $\delta_m > 0$. Fixing the supernova properties, differences in survival probabilities between binary populations can arise due to differences in $v_\mathrm{orb}$ (different companion masses and/or orbital separation) or differences in the post-supernova total binary mass, which affects $\delta_m$. 
}
\end{figure*}

\begin{figure*}
\includegraphics[width=0.9\textwidth]{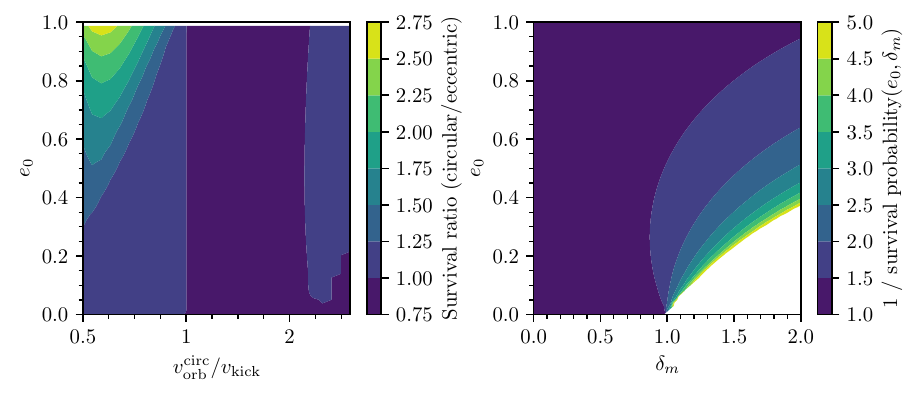}
\caption{\label{fig:ratio_survival_prob_contours_eccentric}
Similar to Fig.~\ref{fig:ratio_survival_prob_contours_dimensionless}, but varying the initial eccentricity $e_0$. The left panel assumes no supernova mass loss ($\delta_m = 0$) and the right panel assumes no natal kicks ($v_\mathrm{kick} = 0$). The left panel fixes $v_\mathrm{orb}^\mathrm{circ}/v_\mathrm{kick}$ to the x-axis, and compares the survival probability of a circular binary to that of an eccentric binary with $e_0$ given by the y-axis. The right panel compares the survival probability of a circular binary with zero mass loss or kick (i.e., survival probability of 1) to that of a binary with mass loss and eccentricity given by ($\delta_m$, $e_0$).
}
\end{figure*}

We explore the hypothesis that kicks are responsible for the dearth of low-mass BHs in both \gaia{} and GW binaries, but that differences in the binary architectures create a deeper gap in \gaia{} compared to GW masses. 
Given a pre-supernova progenitor with mass $m_1^\mathrm{pre}$ in a binary (orbital eccentricity $e_0$ and semimajor axis $a$) with a companion mass $m_2$, the probability that the binary survives a supernova with a randomly oriented (isotropic) natal kick $v_\mathrm{kick}$ that leaves behind a compact object mass $m_1^\mathrm{post}$ is given in terms of dimensionless quantities $v^\mathrm{circ}_\mathrm{orb}/v_\mathrm{kick}$, $\delta_m$ and $e_0$~\citep{1983ApJ...267..322H,1995MNRAS.274..461B,1996ApJ...471..352K}.
The orbital velocity $v_\mathrm{orb}^\mathrm{circ}$ refers to the velocity of one component of the pre-supernova binary relative to the other when their separation is the semi-major axis $a$:
\begin{equation}
    v_{\rm orb}^\mathrm{circ} = \sqrt{G (m_1^{\rm pre} + m_2) / a}.
\end{equation}
We define $\delta_m$ as the dimensionless mass that is lost in the supernova, relative to the post-supernova total mass of the binary:
\begin{equation}
    \delta_m \equiv  \frac{m_1^{\rm pre} - m_1^{\rm post}}{m_1^{\rm post} + m_2}.
\end{equation}
For a given kick velocity $v_\mathrm{kick}$ and supernova mass loss $m_1^\mathrm{pre}-m_1^\mathrm{post}$, differences in eccentricity, semi-major axis, and total mass between \gaia{} and GW systems would lead to different survival probabilities, which may be responsible for the difference in gap depth between the two populations. 
For example, the fact that \gaia{}'s mass gap is a few times emptier than the GW population (see Fig.~\ref{fig:ratio_MG_rate}) could indicate that the corresponding survival probability is a few times smaller. 

In Figs.~\ref{fig:ratio_survival_prob_contours_dimensionless} and~\ref{fig:ratio_survival_prob_contours_eccentric}, we compare the survival probabilities of binaries with different $e_0$, $v_\mathrm{orb}^\mathrm{circ}/v_\mathrm{kick}$ and $\delta_m$.
Each panel of Fig.~\ref{fig:ratio_survival_prob_contours_dimensionless} and~\ref{fig:ratio_survival_prob_contours_eccentric}
shows the ratios of survival probabilities between two binaries with different orbital properties; these ratios of survival probabilities may translate to the ratio of ``mass gap depth" shown in the right panel of Fig.~\ref{fig:ratio_MG_rate}.

Fig.~\ref{fig:ratio_survival_prob_contours_dimensionless} assumes circular pre-supernova binaries ($e_0=0$).
The left panel shows the ratio of survival probabilities between two binaries with different pre-supernova orbital velocities $v_\mathrm{orb}(A)$ and $v_\mathrm{orb}(B)$, assuming both binaries have negligible mass loss compared to the total post-supernova mass ($\delta_m = 0$). 
The survival probability of a given pre-supernova binary approaches zero (one) if its orbital velocity is much smaller (larger) than the kick velocity. When the orbital velocity of binary $A$ is larger than that of binary $B$ (below the diagonal in Fig.~\ref{fig:ratio_survival_prob_contours_dimensionless}, left panel), its survival probability can be several times larger, potentially matching the discrepancy in mass-gap depth between GW and \gaia{} systems, for example. 

Meanwhile, in the right panel of Fig.~\ref{fig:ratio_survival_prob_contours_dimensionless}, we show the effect of the supernova mass loss $\delta_m$ on the ratio of survival probabilities. 
We consider a binary with an arbitrary $\delta_m$ (vertical axis) compared to a binary with $\delta_m = 0$. 
Both binaries have the same pre-supernova orbital velocity $v_\mathrm{orb}/v_\mathrm{kick}$ (horizontal axis). Significant supernova mass loss relative to the post-supernova binary mass decreases the survival probability relative to the negligible mass loss case. 
When the mass loss is sufficiently high, increasing the kick (decreasing $v_\mathrm{orb}/v_\mathrm{kick}$) beyond a certain point actually increases the survival probability of that binary, because fortuitous kicks are needed to keep the binary bound. 

Fig.~\ref{fig:ratio_survival_prob_contours_eccentric} varies the initial eccentricity $e_0$ together with the orbital velocity at separation $a$, $v_\mathrm{orb}^\mathrm{circ}$ (left panel) and the dimensionless supernova mass loss $\delta_m$ (right panel). The left panel assumes negligible supernova mass loss and compares the survival probability for a circular binary with a fixed $v_\mathrm{orb}^\mathrm{circ}/v_\mathrm{kick}$ to an eccentric binary with the same $v_\mathrm{orb}^\mathrm{circ}/v_\mathrm{kick}$. 
If the orbital velocity is small compared to the natal kick, circular binaries are more likely to survive an isotropic natal kick than highly eccentric binaries, by a factor $\lesssim3$. 
If the orbital velocity is large compared to the natal kick and supernova mass loss is negligible, binaries have high survival probabilities regardless of the eccentricity, and eccentric binaries may be slightly more likely to survive.

The right panel of Fig.~\ref{fig:ratio_survival_prob_contours_eccentric} assumes negligible natal kicks ($v_\mathrm{kick} = 0$) and compares the ($\delta_m = 0$, $e_0 = 0$) case (with survival probability of 1) to the ($\delta_m > 0$, $e_0 > 0$) case. If natal kicks are zero, initially circular binaries are unbound due to Blaauw kicks if $\delta_m > 1$, meaning the mass lost in the supernova exceeds the total mass of the post-supernova system. In this regime, increasing the initial eccentricity of the binary increases its survival probability. For $\delta_m > 1$ and moderate initial eccentricities, the survival probability and resulting mass gap will be a few times smaller compared to initially circular binaries with negligible mass loss. 

\todo{Figs.~\ref{fig:ratio_survival_prob_contours_dimensionless} and~\ref{fig:ratio_survival_prob_contours_eccentric} show that differences in pre-supernova orbital velocity (relative to the natal kick), total binary mass (relative to the supernova mass loss), and orbital eccentricity can lead to factor $\mathcal{O}$(few) differences in survival probabilities (and corresponding gap depth). The ratios in survival probabilities between two binary populations can approach infinity if one population has $v_\mathrm{orb} \ll v_\mathrm{kick}$ (e.g., beyond the bottom right corner of~Fig.~\ref{fig:ratio_survival_prob_contours_dimensionless}, left panel) or $\delta_m \gg 1$ (e.g., the top left corner of Fig.~\ref{fig:ratio_survival_prob_contours_dimensionless}, right panel, or the bottom right corner of Fig.~\ref{fig:ratio_survival_prob_contours_eccentric}, right panel). These extremes correspond to one binary population having a survival probability of zero, while the other has a finite survival probability. If supernova natal kicks (or mass loss) are much larger or much smaller than both populations' orbital velocities (or total mass), the ratios of survival probabilities approach unity as both populations have equally small or large survival rates.}

\subsection{Pre-supernova orbital configurations for Gaia and GW binaries and their impact on survival}
The progenitor population to \gaia's mass-gap containing binaries follows a distribution of pre-supernova $v_\mathrm{orb}^\mathrm{circ}$, $\delta_m$, and $e_0$, and this distribution differs from that of GW progenitor binaries. These distributions remain uncertain due to the uncertain formation histories of both populations.
Nevertheless, we can check whether our qualitative expectations produce survival ratios that match the ratio in the mass-gap depth between GW and \gaia{} binaries.
Conversely, if we interpret the ratio in mass gap depths as a ratio in supernova survival probabilities, we may place constraints on the formation histories of the two populations, including the supernovae that produce low-mass BHs.

In the remainder of this section, we explain why \gaia{} progenitor binaries may be less likely to survive a supernova with large mass loss and/or natal kicks than GW progenitor binaries are to survive the same supernovae.
This stems from GW progenitor binaries probably having smaller orbital separations, larger companion masses, and smaller eccentricities than \gaia{} progenitor binaries. 

\subsubsection{Differences in pre-supernova orbital velocities}

Assuming their origin in isolated binary evolution, each \gaia{} NS or BH system has experienced one supernova.
For GW binaries, the supernova producing the mass-gap BH could be either the first or second supernova in the system.
This ambiguity may hold regardless of whether the mass-gap BH is the heavier or lighter component in the binary if mass ratio reversal is common, meaning that the heavier compact object is born first~\citep{2021ApJ...921L...2O, 2022ApJ...933...86Z,2022ApJ...938...45B}. 
In each case, the pre-supernova orbital velocities of progenitor binaries are determined by the ratio of the pre-supernova total mass to the orbital separation. 
\citet{2024A&A...692A.141K} has recently argued that \gaia{} progenitor binaries have wide pre-supernova separations in order to avoid interactions, because binary interactions rarely lead to such extreme mass ratios.\footnote{A small degree of mass transfer (e.g., through wind Roche overflow) may explain the lithium enhancement observed in \gaia{} NS and BH companions through a mechanism similar to the lithium enhancement observed in GALAH red giant population~\citep{2025arXiv250705359S}.} 
They show a representative simulation of a progenitor binary to a \gaia{} BH1 or BH2-like system with a 878.9 day pre-supernova period, corresponding to a $\approx900\,R_\odot$ semi-major axis. Indeed, the orbital separation is larger than the maximum radius of the progenitor star, which can reach a few hundreds of solar radii (or even a thousand solar radii in extreme cases, see their Fig. 1, but note that stellar expansion remains highly uncertain;~\citealt{2022MNRAS.512.5717A,2023MNRAS.525..706R}).
If a mass-gap progenitor binary had a similar orbital separation, the orbital velocity (assuming a circular orbit) is $\sim33$ km/s. 
A significantly smaller orbital separation of $100\,R_\odot$ has a corresponding orbital velocity of $100$ km/s. 
According to supernova simulations, natal kicks accompanying mass-gap BH formation can reach $100$--$1000$ km/s~\citep{2023arXiv231112109B}, so it is reasonable that $v_\mathrm{orb}/v_\mathrm{kick} \lesssim 1$ for \gaia{} binaries. In the regime that $v_\mathrm{orb}/v_\mathrm{kick} < 1$, nonzero orbital eccentricity, which we would expect for wide non-interacting binaries, only decreases the survival probability further (see Fig.~\ref{fig:ratio_survival_prob_contours_eccentric} and the following subsection).

At some point in their evolution, GW binaries necessarily reach much tighter orbital separations than \gaia{} binaries, but the orbital evolution between the first and second supernova is uncertain.
In either case, unlike for \gaia{} wide binaries, the pre-supernova orbits are more likely to be circular due to either mass transfer or tides.

We first consider the case that the mass-gap BH is born in the second supernova. 
Following this second supernova, the periastron separation must be smaller than $\mathcal{O}(10)\,R_\odot$ (depending on the post-supernova eccentricity and masses) so that the two compact objects can merge within a Hubble time~\citep{1964PhRv..136.1224P}.
For example, for a $7\,M_\odot+3\,M_\odot$ BBH with eccentricity $e = 0.7$ (from the supernova), the post-supernova periastron distance (computed according to the~\citealt{2021RNAAS...5..223M} fit to the \citealt{1964PhRv..136.1224P} formula) must be smaller than $6\,R_\odot$.
Assuming an initially circular orbit, the pre-supernova orbital separation will generally be a factor of a few larger than the post-supernova periastron separation~\citep{1983ApJ...267..322H}, or $\lesssim 20\,R_\odot$.

This leads to an orbital velocity that is at least a factor of $\approx3$--$10$ times larger than the \gaia{} progenitor example discussed above, even for the smallest GW companion masses. 
Smaller pre-supernova separations or larger GW companion masses (for example, if the first-born compact object is a $>30\,M_\odot$ BH) can lead to orbital velocities exceeding 1000 km/s, likely in the regime where $v_\mathrm{orb}^\mathrm{circ}/v_\mathrm{kick} > 1$, whereas \gaia{} pre-supernova orbital velocities are reasonably below 100 km/s, with $v_\mathrm{orb}^\mathrm{circ}/v_\mathrm{kick} < 1$.
This can cause the GW survival probabilities at the time of the second supernova to be at least a few times higher than the \gaia{} survival probabilities (see the left panel of Fig.~\ref{fig:ratio_survival_prob_contours_dimensionless}). 

If the mass-gap BH is instead born in the first supernova, the orbital separation is larger, as is the total mass of the system.  
Following the first supernova, the orbital separation needs to be sufficiently small so that stable mass transfer or common envelope can shrink the orbit to below the $\mathcal{O}(10)\,R_\odot$ limit. 
Stable mass transfer may shrink the orbit by factors of 10, while common envelope may be able to shrink it by factors of a few hundred, so that the pre-supernova orbital separations can be $\mathcal{O}(100$--1000) $R_\odot$, similar to or larger than the \gaia{} pre-supernova separations~\citep[see, e.g.,][for a review]{2022PhR...955....1M}. 
However, the companion mass $m_2$ at the time of the first supernova in a GW system is a massive star $m_2\gg10\,M_\odot$ (which will later collapse to a NS or BH), and \gaia{} binaries have companion masses $m_2 \approx 1\,M_\odot$.
Therefore, the binary's total mass following the first supernova is a factor of $\approx10$ times higher for GW than \gaia{} progenitor binaries,
meaning that even accounting for the possibly wider separations of GW progenitors at the time of the first supernova, the orbital velocities will likely be comparable to or larger than the \gaia{} pre-supernova orbital velocities. 
These larger orbital velocities will thus lead to a preference for bound post-supernova GW progenitors and unbound \gaia{} binaries.

\subsubsection{Differences in eccentricity or supernova mass loss relative to binary's total mass}

In the regime where the GW compact object is born with a comparable pre-supernova orbital velocity to that of a \gaia{} progenitor binary, a difference in binary survival probabilities may stem from the pre-supernova orbital eccentricity (see Fig.~\ref{fig:ratio_survival_prob_contours_eccentric}) or supernova mass loss (see the right panel of Fig.~\ref{fig:ratio_survival_prob_contours_dimensionless}). 
In binary evolution, GW pre-supernova binaries likely experienced dissipative interactions (such as tides) that circularized the orbit~\citep{2002MNRAS.329..897H}, while \gaia{} pre-supernova binaries may have avoided interactions~\citep{2024A&A...692A.141K} and follow, e.g., a thermal eccentricity distribution~\citep{1919MNRAS..79..408J}.\footnote{However, not all binary interactions are circularizing; recent work has shown that mass transfer may sometimes increase a binary's eccentricity~\citep{2024A&A...688A.128V,2025ApJ...983...39R}, in line with some observations of post-mass transfer systems~\citep{2017A&A...597A..68V,2018A&A...620A..85O}.}
If the orbital velocity is small compared to the kick velocity for both GW and \gaia{} progenitor binaries, circularized GW binaries would have higher survival probabilities than eccentric \gaia{} binaries by factors of a few (left panel of Fig.~\ref{fig:ratio_survival_prob_contours_eccentric}).

As long as the supernova mass loss is less than $\mathcal{O}(\mathrm{few})\,M_\odot$, it will be negligible compared to the GW binary's total mass ($\delta_m \approx 0$; from the right panel of Fig.~\ref{fig:ratio_survival_prob_contours_dimensionless}, we see that the survival probability changes by less than a factor of 2 for $\delta_m < 0.25$). Meanwhile, for \gaia{} binaries, even $1\,M_\odot$ of mass lost in the supernova is $\approx25\%$ of the total mass, such that any more supernova mass loss could decrease the survival probability by more than a factor of 2, depending on $v_\mathrm{orb}^\mathrm{circ}/v_\mathrm{kick}$.  
This order-of-magnitude difference in $\delta_m$ at the time of the first supernova may contribute to the preferential survival of GW binaries containing first-born mass-gap BHs.

\section{Conclusion}
\label{sec:conclusion}

In this work, we compared the component mass distribution of GW systems to the mass distribution of NS and BH in wide \gaia{} binaries, focusing on the transition between NS and BH masses (i.e. the ``mass gap" region). 
We then discussed the possible role of supernova kicks in shaping the different occurrence rates of low-mass BHs in the difference systems. We found that:
\begin{itemize}

\item Current observations suggest that low-mass BHs in the ``mass gap" are relatively more common in GW binaries than \gaia{} binaries at 94\% credibility, or 76\% credibility if including G3425; see Fig.~\ref{fig:Gaia_versus_LVK_massdist}. The fraction of BHs with masses between 2.5 and 5 $M_\odot$ is a factor of $11^{+1385}_{-10}$ higher in GW binaries than \gaia{} binaries (or $1.9^{+12.4}_{-1.5}$ if we include the \gaia{} mass-gap candidate G3425); see Fig.~\ref{fig:ratio_MG_rate}.

\item Before the second supernova, GW progenitor binaries are likely an order of magnitude tighter (in orbital separation) than \gaia{} progenitor binaries before the first and only supernova, implying that the pre-supernova orbital velocity is at least an order of magnitude larger ($\mathcal{O}(100)$ km/s compared to $\mathcal{O}(10)$ km/s). If the mass-gap BH is born in the second supernova of a GW progenitor, and the supernova imparts a kick of $\mathcal{O}(100)$ km/s, the GW progenitor binary will be a few times more likely to survive than a \gaia{} progenitor binary; see the left panel of Fig.~\ref{fig:ratio_survival_prob_contours_dimensionless}.

\item Following the first supernova, GW binaries are likely an order of magnitude heavier than \gaia{} binaries. If the mass-gap BH is born in the first supernova of a GW progenitor and the mass lost in the supernova is $\mathcal{O}(1)\,M_\odot$, the mass loss will have a negligible effect on the binary survival probability. On the other hand, the supernova mass loss could decrease the survival probability of \gaia{} progenitor binaries by at least a factor of 2; see the right panel of Fig.~\ref{fig:ratio_survival_prob_contours_dimensionless}. Pre-supernova orbital eccentricity can also decrease the survival probability of \gaia{} binaries, compared to GW binaries which are circular before each supernova (Fig.~\ref{fig:ratio_survival_prob_contours_eccentric}).
\end{itemize}

Therefore, the natal kicks and mass loss accompanying the birth of low-mass BHs could credibly lead to a higher relative rate of mass-gap BHs among GW binaries compared to \gaia{} binaries, matching the observed discrepancy in the depth of their mass gaps. Of course, there are other potential explanations as well, involving environmental differences (e.g., metallicity), binary physics, or evolutionary pathways not discussed here, such as the effect of a triple companion or dynamical assembly in clusters~\citep[e.g.,][]{2023MNRAS.526..740R,2024ApJ...965...22D,2024A&A...688L...2M,2024MNRAS.527.4031T,2024MNRAS.528.5119A,2025arXiv250616513S}.
Although a detailed analysis of these other evolutionary effects is beyond the scope of this work, we note that natal kicks may still shape the mass gap in the context of stellar dynamics. 
Indeed, \gaia{} BH3 -- which we do not consider in our analysis -- likely originated in a low-mass star cluster since its present day location has been localized to the GD-1 stellar stream~\citep{2024A&A...687L...3B,2025arXiv251007021M}.
If \gaia{} compact objects originate in smaller clusters (with smaller escape speeds) than GW compact objects, the BHs that receive large natal kicks will be preferentially ejected from their clusters, making them less likely to be observed as \gaia{} binaries~\citep{2024OJAp....7E..39T}. Thus, a possible difference in cluster escape speeds may similarly lead to a deeper mass gap among the \gaia{} population compared to the GW population. 

With the current sample size of low-mass BHs, there are large uncertainties in the depth of the mass gap among GW and \gaia{} binaries. 
Future detections of GW sources and wide astrometric binaries, together with observations of X-ray binaries~\citep[e.g.][]{2023MNRAS.525.1498Z} and isolated NSs and BHs via microlensing~\citep[e.g.][]{2020ApJ...889...31L,2022ApJ...930..159A}, will yield improved population statistics, refining our understanding of the ``mass gap" region in the different populations. 
\todo{If the tension in mass gap depth between the \gaia{} and GW population disappears with future observations, it may indicate that supernova kicks received by low-mass BHs are either very large or very small, washing out differences in the orbital architectures of the two populations.}

Future observations will allow us to further resolve the properties of mass-gap events. Upcoming data releases from \gaia{} will probe binaries with shorter orbital periods and higher companion masses that were not accessible in the latest data release \citep{Holl+2023:2023A&A...674A..25H}, which may reveal a negative correlation between the rate of low-mass BHs and binary separation, or a positive correlation with companion mass, suggesting the influence of supernova kicks on binary survival.
Larger GW catalogs will allow us to distinguish between the mass distributions of first-born versus second-born BHs~\citep[e.g.][]{2024ApJ...962...69F}. We expect the mass-gap rate to differ between first-born and second-born BHs if supernova kicks are responsible for the gap, because the progenitor binaries have different architectures between the first and second supernova. If supernova natal kicks are primarily responsible for binary disruption, we expect a deeper mass gap among first-born BHs, whereas if mass loss is primarily responsible, the mass gap may be deeper among second-born BHs.  

The statistics of the mass gap in different populations, combined with
direct measurements of the kicks of survived binaries, using peculiar velocities, post-supernova orbital properties, and the spin tilts of GW sources, will allow us to simultaneously constrain supernova properties and the evolutionary histories of NSs and BHs in different systems.
Such studies highlight the power of asynchronous multi-messenger astronomy in probing stellar evolution, binary physics and the origin of NSs and BHs.


\bibliography{references}{}
\bibliographystyle{aasjournal}

\section*{Statements and Declarations}

\centerline{\bf Acknowledgments}
We thank Alexandra Guerrero for comments and suggestions on the manuscript. We are grateful to the Lorentz Center, the organizers and participants of the workshops ``Gravitational waves: a new ear on the chemistry of galaxies" and ``Challenges and future perspectives in gravitational-wave astronomy," for inspiring this work. This material is based upon work supported by NSF's LIGO Laboratory which is a major facility fully funded by the National Science Foundation.
\vspace{20pt}

\centerline{\bf Funding}
MF acknowledges support from the Natural Sciences and Engineering Research Council of Canada (NSERC) under grant RGPIN-2023-05511, the University of Toronto Connaught Fund, the Alfred P. Sloan Foundation, and the Ontario Early Researcher Award. 
RW acknowledges support from the KU Leuven Research Council through grant iBOF/21/084.  
\vspace{20pt}

\centerline{\bf Competing Interests}
The authors have no relevant financial or non-financial interests to disclose.

\vspace{20pt}

\centerline{\bf Ethics Declaration}
Not applicable.

\end{document}